\documentclass[9pt,twocolumn,twoside]{opticajnl}

\journal{ol} 

\setboolean{shortarticle}{true}

\usepackage{lineno}
\usepackage{subfig}

\title{A monolithic interferometer for high-sensitive strictly-local detection of orbital angular momentum states of light}

\author[1,*]{Mirko Siano}
\author[1]{Bruno Paroli}
\author[1]{Simone Cialdi}
\author[1]{Stefano Olivares}
\author[1]{Matteo G. A. Paris}
\author[1]{Edoardo Suerra}
\author[1]{Marco A. C. Potenza}

\affil[1]{Dipartimento di Fisica, Università degli Studi di Milano and INFN Sezione di Milano, via G. Celoria 16, 20133 Milan, Italy}

\affil[*]{Corresponding author: mirko.siano@unimi.it}

\begin{abstract}

We propose an innovative monolithic interferometer to distinguish the topological charge of radiation carrying orbital angular momentum. Remarkably, our method requires to access only a small portion of the entire wavefront. The proposed scheme relies on a monolithic birefringent crystal, and as such it is intrinsically stable and does not require any feedback or thermal drift compensation. An experimental setup has been realized to prove the effectiveness of the proposed method down to the photon counting regime.
\end{abstract}

\setboolean{displaycopyright}{true}

\begin{document}

\maketitle


Radiation with Orbital Angular Momentum (OAM) formalized at the end of the twentieth century~\cite{oam} has in recent decades aroused considerable interest in various scientific disciplines~\cite{resolution, transitions, astro, telecomm1, telecomm2, nanoparticles, bpms_dense_code}, effectively leading to the formation of what is now known as singular optics~\cite{singular_optics}. 

Common to the various disciplines and applications, detection of OAM radiation at the level of single photon in high-sensitive detectors is of particular interest in both classical~\cite{oam_classic_1, oam_classic_2} and quantum realms~\cite{oam_quantum_1, oam_quantum_2}. Another fundamental aspect is the ability of measuring the topological charge using only a small portion of the propagating beam~\cite{bpms_dense_code, bpms_ao, bpms_composite, bpms_hybrid}. Recognizing OAM states in a completely local manner would open important perspectives, not yet explored, for a novel class of detectors and receivers. However, achieving high-sensitive detection of OAM states in a strictly local manner is still a challenge, as generally the entire wavefront of radiation is necessary both with refractive~\cite{refractive1} and diffractive~\cite{diffractive1, diffractive2} state-of-the-art techniques. 

In this Letter, we report for the first time a novel high-sensitive local detection method that exploits only a small portion of the beam (also far from the singularity) to distinguish different OAM states. We also prove that the proposed scheme performs extremely well and reliably even with a very small collection of photons.

The method exploits the peculiar phase properties of OAM radiation in a two-point detection scheme. Let $x-y$ be a plane transverse to the propagation direction of an OAM beam. The phase $\phi$ of the associated electric field varies as a function of the azimuthal angle $\theta$ on the $x-y$ plane according to $\phi = l \theta$, where $l$ is the topological charge. Therefore, a measurement of the phase difference $\Delta \phi=\phi_2-\phi_1$ at two different angular positions $\theta_1$ and $\theta_2$ separated by $\Delta \theta=\theta_2-\theta_1$ provides the OAM state $l=\Delta \phi/\Delta \theta$. Here we show how to measure the phase difference $\Delta \phi$ with an extremely robust and strictly local monolithic interferometer based on a birefringent calcite crystal. The proposed scheme exhibits high resilience to external perturbations and enables working in photon counting regime for long times without the need of any drift compensation, an essential requirement for high-sensitive detection.

We exploit the birefringent nature of the calcite crystal to realize a monolithic wavefront-division interferometer endowed with extreme stability and resistance to external perturbations, according to the following scheme. We consider two points $P_1$ and $P_2$ on the front face of the crystal, where $P_1$ and $P_2$ represent the input ports of the interferometer. They are at the same radial distance $r$ from the OAM singularity and are separated by $\Delta x$, thus they subtend the angle $\Delta\theta \sim \Delta x/r$ with respect to the center of the OAM beam. We then cover either $P_1$ or $P_2$ with a $\lambda/2$ wave plate oriented at 45 degrees with respect to the polarization axis of the incoming OAM beam, which is made coincident with the polarization axis of the ordinary ray of the crystal. Thanks to this configuration, the electric fields at $P_1$ and $P_2$ have orthogonal polarizations and propagate through the crystal along the ordinary and extraordinary paths, respectively, and superimpose on an observation plane downstream the crystal. By projecting both fields along a common polarization direction with a linear polarizer, interference fringes form:

\begin{equation}\label{intensity}
I(x)= 4 I_0 \frac{\cos{(k_f x+l \Delta \theta+\phi_0)}+1}{2},
\end{equation}

\noindent where $I_0$ is the intensity of the incident beam, $k_f=2 \pi/\lambda_f$, $\lambda_f$ being the spatial period of the interference fringes, and $\phi_0$ is the phase shift due to the difference between the ordinary and extraordinary paths. In Eq.~\eqref{intensity} we assume, for simplicity, that fringes are oriented along the vertical direction $y$, with $x$ the horizontal axis.

\begin{figure*}
\centering\includegraphics[width=\textwidth]{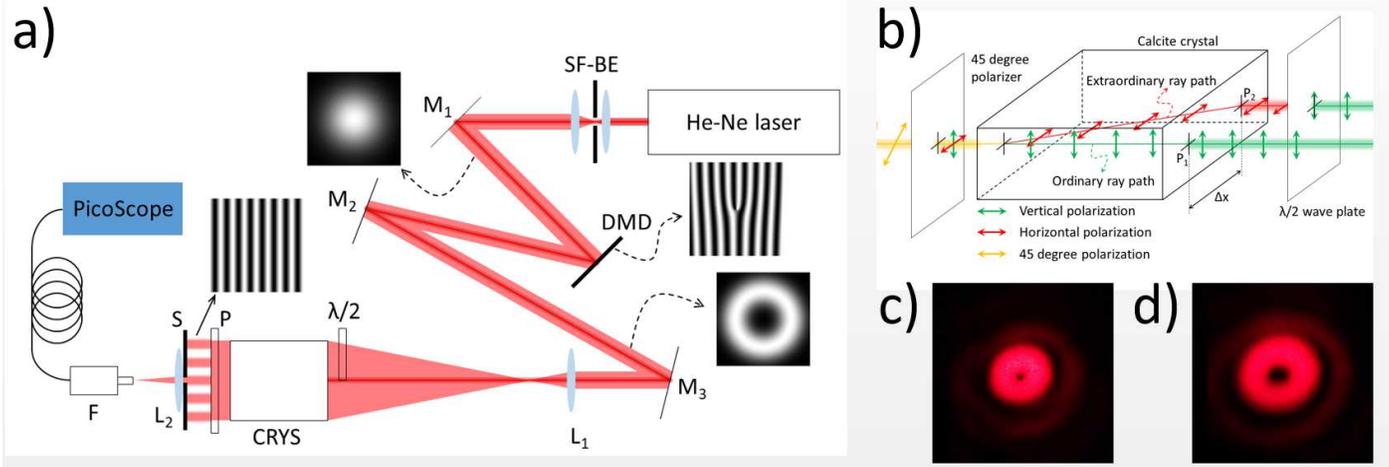}
\caption{(a) Experimental setup. SF-BE: spatial filter and beam expander; M1, M2, M3: mirrors; DMD: Digital Micromirror Device; L1, L2: positive lenses; $\lambda/2$: half-wavelength wave plate; CRYS: birefringent calcite crystal; P: 45 degrees polarizer; S: narrow slit; F: single-mode optical fiber coupled to L2 and to a fast acquisition system PicoScope. (b) Detail of the monolithic interferometer based on the birefringent calcite crystal. The case of a perfectly collimated incident beam is depicted. (c) and (d) Intensity profiles of the OAM beams with $l=1$ and $l=2$, respectively, at the entrance of the monolithic interferometer.}
\label{setup}
\end{figure*}

Notice that the dependence on the topological charge in Eq.~\eqref{intensity} is given only by the linear term $l \Delta \theta$, since $k_f$ and $\phi_0$ are independent on $l$. Therefore, the OAM state can be directly detected either from the absolute position of the fringes, or from their relative intensity values at a particular observation point $x$.

\begin{figure*}
\centering\includegraphics[scale=0.6]{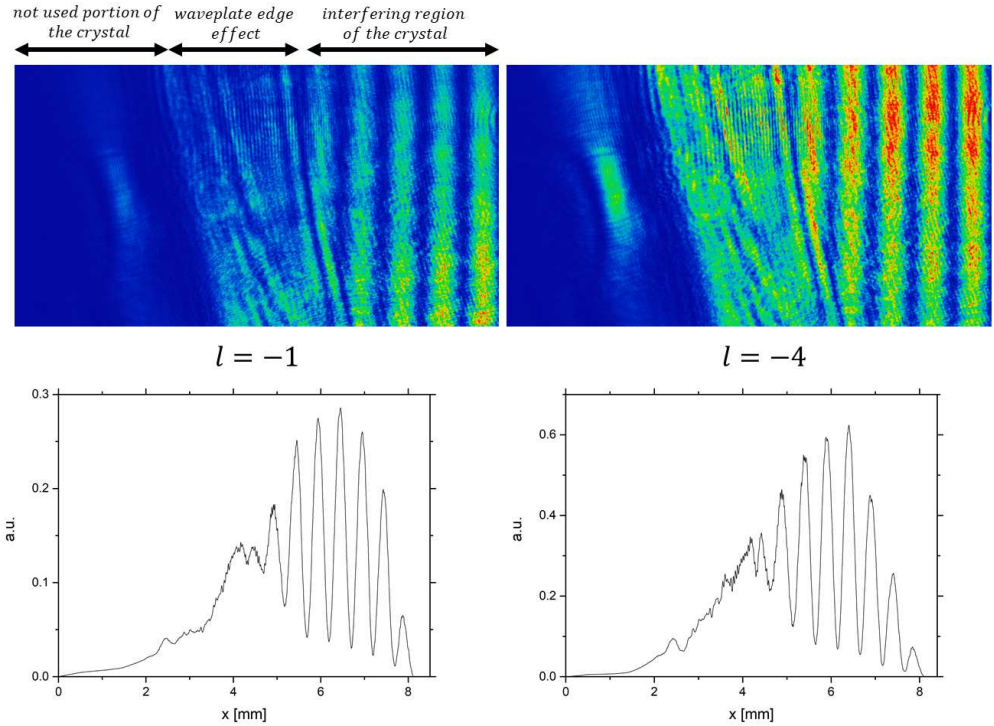}
\caption{Top: Portion of the interference pattern of the monolithic interferometer for the topological charges $l=-1$ and $l=-4$. Interference fringes generated by the superposition of the beams propagating along the ordinary and extraordinary paths are clearly visible to the right. Bottom: Intensity profiles of the interference patterns integrated along the fringe direction.}
\label{patterns}
\end{figure*}

A proof-of-principle setup was realized to prove the effectiveness of the novel local detection method as sketched in Fig.~\ref{setup} (a). A He-Ne laser beam with $\lambda$=632.8 nm is spatially filtered and collimated. The resulting Gaussian beam has a diameter of 3.7 mm (1/$e^2$ intensity value) and illuminates a Digital Micromirror Device (DMD). A computer-generated fork hologram is projected onto the DMD to convert the Gaussian beam into an OAM beam. The first diffraction order is selected, corresponding to a OAM beam with the desired topological charge. The OAM beam is then properly aligned along the horizontal and vertical planes and sent to the interferometer by means of two mirrors. The beam is expanded through a positive lens before impinging onto a birefringent calcite crystal (Thorlabs BD40). The beam diameter at the entrance of the crystal is roughly 33 mm for an $l=0$ OAM state corresponding to a standard Gaussian beam (1/$e^2$ intensity value). Half of the birefringent crystal is covered by a $\lambda/2$ wave plate, oriented at 45 degrees with respect to the vertical polarization of the incoming OAM beam.  This ensures that the polarization of the transmitted portion of the OAM wavefront is rotated along the horizontal plane. The vertically and horizontally polarized portions of the incoming OAM beam propagate inside the calcite crystal along the directions parallel to the ordinary and extraordinary rays, respectively, and superimpose downstream the crystal. A 45 degrees polarizer ensures interference between these two wavefronts. Interference fringes are acquired with a standard CMOS camera (Thorlabs DCC1240M, 1280 x 1024 pixels, 5.3 x 5.3 \textmu m$^2$ pixel size).

A detailed view of the monolithic interferometer is sketched in Fig.~\ref{setup} (b). For the case of a perfectly collimated incident beam, the ordinary and extraordinary rays which, at the entrance of the crystal, are separated by a suitable $\Delta x$, exit the crystal from the same point and propagate collinearly and perfectly overlapped along the same direction, giving rise to a uniform intensity distribution upon interference ($\lambda_f \rightarrow \infty$ in Eq.~\eqref{intensity}). The particular value of $\Delta x$ for which this occurs depends on the crystal and the radiation wavelength. It is roughly 4.3 mm in our case.

Contrarily, if the incident wavefront is endowed with a finite curvature, the two rays emerge from the crystal from two different points and superimpose on the observation plane with a finite angle $\Theta$, thus forming interference fringes with finite periodicity $\lambda_f \sim \lambda/\Theta$. In such case, the two rays originate from portions of the incoming wavefront whose separation $\Delta x$ is smaller compared to the case of a perfectly collimated beam, the exact value depending on the actual wavefront curvature. In our case, for a radius of curvature of 2.6 m, the actual separation between the two points $P_1$ and $P_2$ is 3.3 mm.



The interference patterns acquired with the CMOS camera at the exit of the monolithic interferometer are shown in Fig.~\ref{patterns} (top) for $l=-1$ and $l=-4$. They are numerically integrated along the direction of the interference fringes to extract the interference profiles shown in Fig.~\ref{patterns} (bottom), while at the same time reducing noise. Notice that the rightmost portion of the pattern is characterized by the expected interference fringes due to the superposition of the ordinary and extraordinary rays from the crystal (projected along a common polarization direction), while the leftmost portion shows distortions introduced by diffraction at the edge of the $\lambda/2$ wave plate.

\begin{figure*}
\centering
\subfloat
{\includegraphics[width=.45\textwidth]{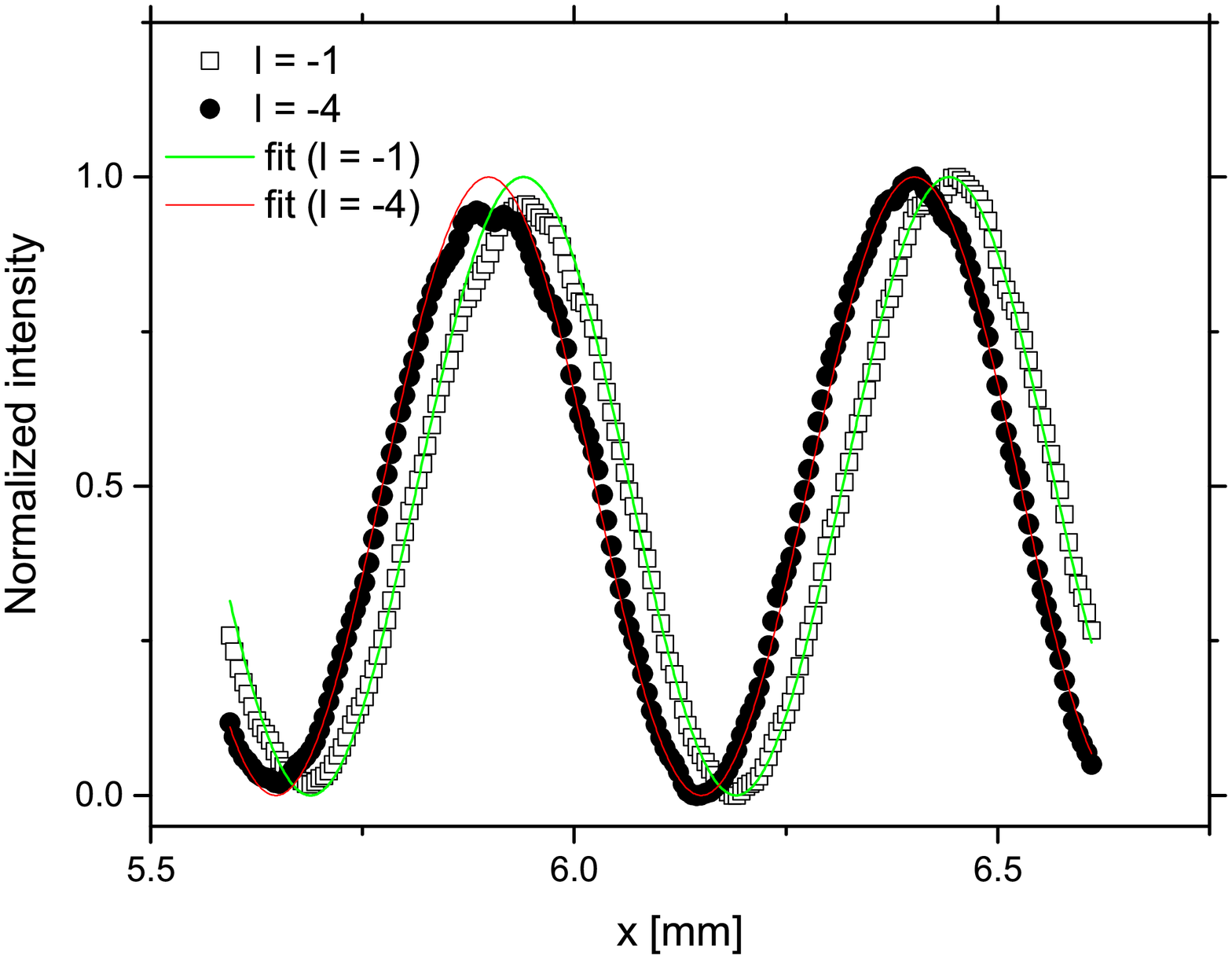}}
\subfloat
{\includegraphics[width=.45\textwidth]{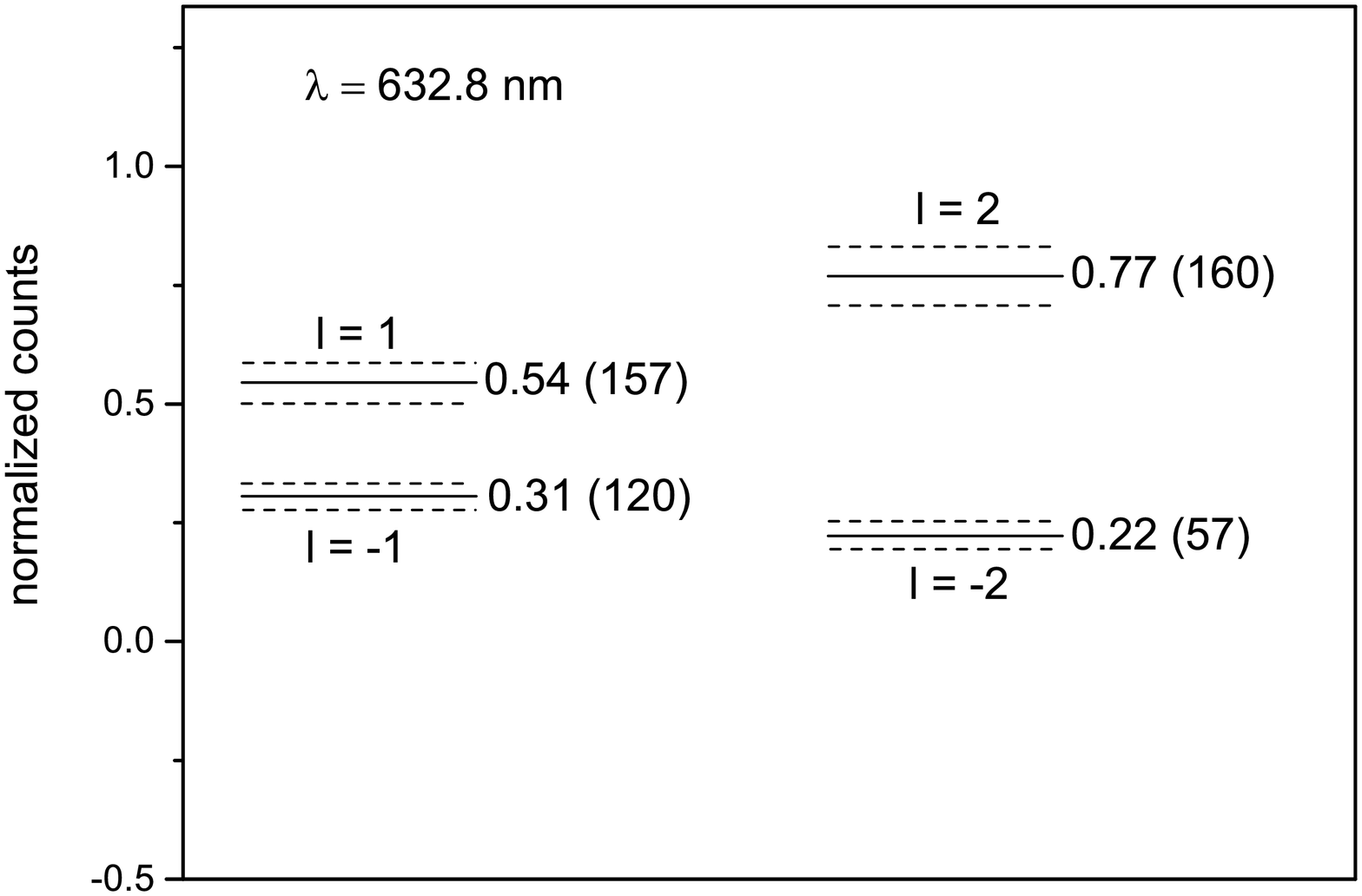}}
\caption{Left: Normalized intensity $I_n$ of the integrated interference pattern generated with the monolithic interferometer for the OAM states $l=-1$, $l=-4$. The red and green curves represent fits to Eq.~\eqref{intensity} to extract the phase parameters. Right: Normalized counts obtained with the monolithic interferometer working in photon counting mode for different OAM states. Dashed lines show the statistical errors. Absolute photon counts are shown in brackets.}
\label{phase}
\end{figure*}

To check the effectiveness of the interferometer, we precisely measure the phase shift of the fringes upon changing the topological charge from $l=-1$ to $l=-4$. We report in Fig.~\ref{phase} (left) the same integrated intensity profiles as in Fig.~\ref{patterns} (bottom) in the range $5.6-6.6$ mm, properly normalized according to 

\begin{equation}\label{normalization}
I_n=\frac{I-\text{min}\{I\}}{\text{max}\{I-\text{min}\{I\}\}},
\end{equation}

\noindent where $I$ is the intensity profiles in Fig.~\ref{patterns} (bottom) and $I_n$ is the corresponding normalized version. Normalization according to Eq.~\eqref{normalization} allows to distinguish the OAM states from the position of the fringes despite the absolute intensity values of the $l=-1$ and $l=-4$ OAM beams substantially changes. The measured phase shift is $0.513 \pm 0.007$ rad, in excellent agreement with the calculated value $0.47\pm 0.05$ rad. The main source of uncertainty for the calculated value is the position of the local interferometer with respect to the singularity ($r=21 \pm 2$ mm). The experimental phase shift was evaluated using the phase parameters extracted by a least squares method to Eq.~\eqref{intensity} and the experimental error stems from the fit uncertainty.

To test the performances of the interferometer in detecting OAM states with high-sensitivity, we push the system into the photon counting regime. To this aim, a small portion of the interference pattern is selected through a narrow slit with aperture 50 \textmu m, much smaller than the fringe period. The light transmitted by the slit is collected by a single-mode fiber coupled to a positive lens. The extremely small aperture of the fiber (roughly 8 \textmu m) and the tight alignment requirements for proper lens-fiber coupling ensure the photon counting regime. The resulting photon counts are read by a photomultiplier and converted to electric pulses measured by a fast acquisition card (Picoscope 4224A).

We adjust the lateral position of the slit across the interference plane to properly set the working point of the detector close to the region of the interference pattern with the highest intensity derivative, to maximize sensibility~\cite{bpms_hybrid}. Notice that this is made possible by the small, albeit finite wavefront curvature imposed by the positive lens before the interferometer, which results in interference fringes with a finite periodicity. Contrarily, for a perfectly collimated OAM beam with vanishing curvature, a uniform intensity pattern would form downstream the crystal, as previously discussed. Therefore, based on Eq.~\eqref{intensity}, we optimize the states separation by experimentally setting the working point of the interferometer to $k_f x+\phi_0 \approx \pi (m+1/2)$, where $m$ is an integer~\cite{bpms_hybrid}. We do so by sending an $l=0$ state (corresponding to a standard Gaussian beam) to the interferometer, and by adjusting the slit position halfway between the maximum and the minimum intensity of an interference fringe~\cite{bpms_hybrid}.

After this initial calibration, counts relative to the different OAM states $l=-1$, $l=1$, $l=-2$, $l=2$ have been measured, each in an interval time of $100$ ms. For each OAM state, counts are then normalized to their maximum value, which is obtained when the slit is positioned in correspondence of an intensity maximum of the fringe pattern. This enables direct comparison among the different OAM states, regardless of the overall intensity level $I_0$ in Eq.~\eqref{intensity} for the different OAM beams, similarly to the procedure adopted in Eq.~\eqref{normalization} and below it. Results are shown in Fig.~\ref{phase} (right). The four OAM states are well distinguishable by using $\leq$ 160 photon counts only. Notice that the splitting of the states $l=\pm 2$ is twice as much that of the states $l\pm 1$, as predicted by theory.

Finally, to quantify the stability of the interferometer, we monitored any eventual drift of the interference fringes generated by either the $l=-1$ or the $l=-4$ OAM beam over long acquisitions. In the worst case, we measured a maximum phase change of less than 0.25 rad over more than one hour, corresponding to a phase drift of 70 \textmu rad/s, likely in presence of mechanical or thermal relaxations. The impact on our measurements is definitely negligible, since this maximum drift would introduce a phase shift 30 times smaller than the measured phase change of 0.513 rad between the two images acquired with a temporal delay of 240 s in Fig.~\ref{phase} (left), and orders of magnitude smaller than the phase changes between either the $l=\pm1$ states or the $l=\pm2$ states during the photon counting experiments, with acquisitions being performed over a total time of 200 ms. This proves that the monolithic interferometer fulfills the requirements of stability during normal operations, and does not require any thermal drift compensation to distinguish different OAM states over minutes-long acquisitions.


In conclusions, we have shown a novel, intrinsically stable monolithic interferometer based on a birefringent calcite crystal, capable of detecting the topological charge of radiation carrying orbital angular momentum by exploiting a very small portion $f=5\%$ of the entire wavefront only ($f=2\Delta x L / \pi r^2$, $L=11$ mm being the vertical size of the calcite crystal). An experimental setup has been realized to prove the effectiveness of the proposed method through the absolute measurement of the phase shift introduced by changing the OAM states. Experimental results are in good agreement with theory. Furthermore, the high stability of the interferometer allows photon counting measurements of OAM states. Results prove that the states $l=-1$, $l=1$, $l=-2$ and $l=2$ can be detected with only a few hundreds of photon counts, opening new perspectives for high-sensitive local measurements of OAM radiation without the need of thermal drifts compensation and with high resilience to mechanical vibrations.

In the current setup, the photon counting experiment has been conducted by operating the monolithic interferometer in combination with a single slit. Therefore, the sensibility can be further increased by employing e.g. a periodic array of slits, or a grating, having the same fringe periodicity. Acting on several fringes simultaneously, the overall accuracy and sensibility can then be greatly enhanced. This is not necessary when the wavefront curvature approaches zero, since the light intensity concentrates in a single fringe downstream the crystal exit port.

The proposed method can be advantageously applied to detect OAM photon states at large distances from the source, where it is not possible to access the entire wavefront of the radiation beam, or in most applications where the position of the singularity cannot be precisely aligned, especially in long-distance free-space communications. Our findings can also foster the development of new detection schemes in the quantum regime.

\begin{backmatter}

\bmsection{Acknowledgments} This work was supported by the National Institute for Nuclear Physics (INFN) (project: ADAMANT).

\smallskip

\bmsection{Disclosures} The authors declare no conflicts of interest.

\bmsection{Data availability} Data underlying the results presented in this paper are not publicly available at this time but may be obtained from the authors upon reasonable request.

\end{backmatter}


\bibliography{sample}

\bibliographyfullrefs{sample}

\end{document}